\documentclass[final,5p,twocolumn]{elsarticle}

\usepackage{amssymb}
\usepackage{amsmath}
\usepackage{hyperref}

\newcommand{\dd}{\mbox{$\textrm{d}$}}

\journal{Physics Letters B}

\begin{document}

\begin{frontmatter}

\title{Experimental and theoretical study of deuteron-proton elastic scattering for\\ proton kinetic energies between $T_p = 882.2\;\textrm{MeV}$ and $T_p = 918.3\;\textrm{MeV}$}

\address[munster]{Institut f\"ur Kernphysik, Westf\"alische Wilhelms-Universit\"at M\"unster, Wilhelm-Klemm-Str. 9, 48149 M\"unster, Germany}
\address[gatchina]{High Energy Physics Department, St. Petersburg Nuclear Physics Institute, RU-188350 Gatchina, Russia}
\address[julich]{Institut f\"ur Kernphysik, Forschungszentrum J\"ulich, D-52425 J\"ulich, Germany}
\address[dubna]{Laboratory of Nuclear Problems, JINR, RU-141980 Dubna, Russia}
\address[cracow]{H. Niewodniczanski Institute of Nuclear Physics PAN, PL-31342 Cracow, Poland}
\address[moscow]{Skobeltsyn Institute of Nuclear Physics, Lomonosov Moscow State University, Leninskie Gory 1/2, 119991 Moscow, Russia}
\address[tiblis]{High Energy Physics Institute, Tbilisi State University, GE-0186
	Tbilisi, Georgiam}
\address[london]{Physics and Astronomy Department, UCL, Gower Street, London WC1E 6BT, United Kingdom}

\author[munster]{C.~Fritzsch\corref{mycorrespondingauthor}}
\cortext[mycorrespondingauthor]{Corresponding author}
\ead{c.fritzsch@uni-muenster.de}

\author[gatchina]{S.~Barsov}
\author[munster]{I.~Burmeister}
\author[julich,dubna]{S.~Dymov}
\author[julich]{R.~Gebel}
\author[julich]{M.~Hartmann}
\author[julich]{A.~Kacharava}
\author[munster]{A.~Khoukaz}
\author[dubna]{V.~Komarov}
\author[cracow]{P.~Kulessa}
\author[dubna]{A.~Kulikov}
\author[moscow]{V.~I.~Kukulin}
\author[julich]{A.~Lehrach}
\author[julich]{B.~Lorentz}
\author[julich,tiblis]{D.~Mchedlishvili}
\author[munster]{T.~Mersmann}
\author[munster]{M.~Mielke}
\author[gatchina]{S.~Mikirtychiants}
\author[julich]{H.~Ohm}
\author[munster]{M.~Papenbrock}
\author[moscow]{M.~N.~Platonova}
\author[julich]{D.~Prasuhn}
\author[julich]{V.~Serdyuk}
\author[julich]{H.~Str\"oher}
\author[munster]{A.~T\"aschner}
\author[julich,gatchina]{Yu.~Valdau}
\author[london]{C.~Wilkin}

\begin{abstract}
New precise unpolarised differential cross sections of deuteron-proton
elastic scattering have been measured at 16 different deuteron beam momenta
between $p_d = 3120.17\;\textrm{MeV}/c$ and $p_d =3204.16\;\textrm{MeV}/c$ at
the COoler SYnchrotron COSY of the Forschungszentrum J\"ulich. The data,
which were taken using the magnetic spectrometer ANKE, cover the equivalent
range in proton kinetic energies from $T_p = 882.2\;\textrm{MeV}$ to $T_p =
918.3\;\textrm{MeV}$. The experimental results are analysed theoretically
using the Glauber diffraction model with accurate nucleon-nucleon input. The
theoretical cross section at $T_p = 900\;\textrm{MeV}$ agrees very well with
the experimental one at low momentum transfers $|t| <
0.2\;(\textrm{GeV}/c)^2$.
\end{abstract}


\end{frontmatter}

\section{Introduction}
\label{Intro}

\begin{figure}[h]
\centering
\includegraphics[width=1\linewidth]{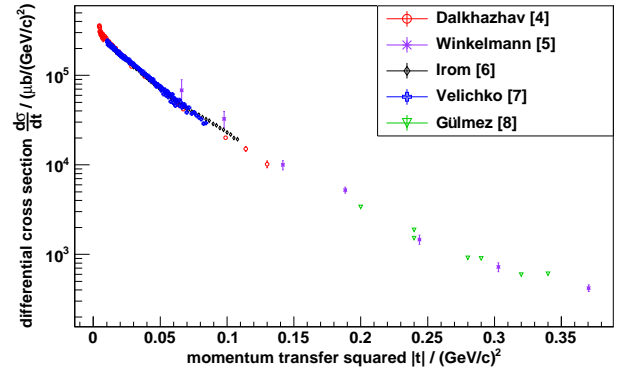}
\caption{Unpolarised differential cross sections of $dp$ elastic scattering
plotted as a function of the momentum transfer squared $-t$ for different
data
sets~\cite{Dalkhazhav:1969cma,Winkelmann:1980ca,Irom:1984wr,Velichko:1988ed,Guelmez:1991he}.}
\label{fig:dpreferencedatalog}
\end{figure}

Deuteron-proton elastic scattering is extensively used in the study of, e.g.,
meson production mechanisms in few nucleon systems at intermediate energies.
For such experiments $dp$ elastic scattering is well suited for normalisation
purposes, due to its high cross section over a large momentum transfer range
(cf.\ Fig. \ref{fig:dpreferencedatalog}). Previous work on meson production,
e.g., Refs. \cite{Mersmann:2007gw,Rausmann:2009dn,Mielke:2014xbu}, used the
existing
database~\cite{Dalkhazhav:1969cma,Winkelmann:1980ca,Irom:1984wr,Velichko:1988ed,Guelmez:1991he}
for data normalisation, assuming that for low momentum transfers, i.e., $|t|
< 0.4\;(\textrm{GeV}/c)^2$, the differential cross section as a function of
$t$ is independent of the beam momentum in the proton kinetic energy range
between $T_p = 641\;\textrm{MeV}$ and
$T_p = 1000\;\textrm{MeV}$. 

In contrast to the database at smaller momentum transfers $|t| <
0.1\;(\textrm{GeV}/c)^2$, that at larger $|t|$ is much poorer. High-precision
data from the ANKE spectrometer, using a deuteron beam and a hydrogen target,
allows further study of the behaviour of the unpolarised differential cross
sections. This enlarges the database in the momentum transfer range $0.08 <
|t| < 0.26\;(\textrm{GeV}/c)^2$ at deuteron momenta that correspond to proton
energies between $T_p = 882.2\;\textrm{MeV}$ and $T_p = 918.3\;\textrm{MeV}$.

On the theoretical side, $pd$ elastic scattering in the GeV energy region has
usually been analysed in terms of the Glauber diffraction model (or its
various extensions), which is a high-energy and low-momentum-transfer
approximation to the exact multiple-scattering series for the hadron-nucleus
scattering amplitude. The original Glauber model~\cite{Franco:1965wi}, where
spin degrees of freedom were neglected (or included only partially), has been
refined~\cite{Platonova:2016xjq,Platonova:2010zz} by taking fully into
account the spin structure of colliding particles, i.e., the spin-dependent
$NN$ amplitudes and the $D$-wave component of the deuteron wave function, and
also the double-charge-exchange process $p+d \to n+(pp) \to p+d$. In
addition, while the majority of previous calculations made within the Glauber
model employed simple parameterisations for the forward $NN$ amplitudes, the
refined model~\cite{Platonova:2016xjq,Platonova:2010zz} suggests using
accurate $NN$ amplitudes, based on modern $NN$ partial-wave analysis (PWA).
By using the $NN$ PWA of the George Washington University SAID group
(SAID)~\cite{Arndt:2007qn}, the model has been shown to describe small-angle
$pd$ differential cross sections and also the more sensitive polarisation
observables very well in the energy range $T_p =
200$\,--$1000\;\textrm{MeV}$~\cite{Platonova:2016xjq}. The refined Glauber
model therefore seems ideally suited for the description of the experimental
data presented here. On the other hand, the new high-precision data can
provide a precise test for applicability of the Glauber model.

The SAID group has recently published an updated $NN$ PWA
solution~\cite{Workman:2016ysf}, which incorporates the new COSY-ANKE data on
the near-forward cross section~\cite{Mchedlishvili:2015iwa} and analysing
power $A_y$~\cite{Bagdasarian:2014mdj} in $pp$ elastic scattering, as well as
the recent COSY-WASA $A_y$ data~\cite{Adlarson:2014pxj} in $np$ elastic
scattering. We can therefore re-examine the predictions of the refined
Glauber model obtained with the use of the previous PWA solution of
2007~\cite{Arndt:2007qn}. By performing calculations at various incident
energies, we can also test the widely-used assumption of energy independence
of the $pd$ elastic differential cross section at low momentum transfers.

\section{Experimental Setup}
\label{ExpSetup}

\begin{figure}[!ht]
\centering
\includegraphics[width=1\linewidth]{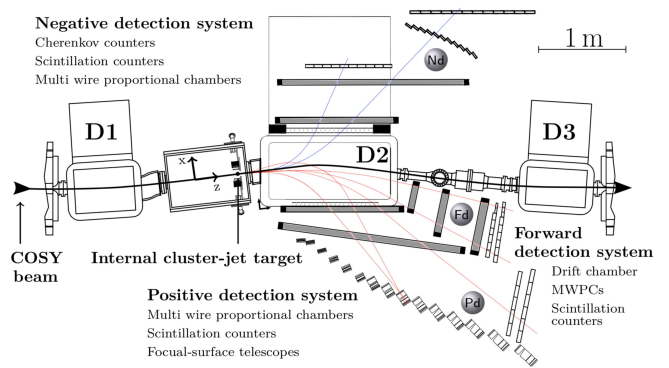}
\caption{Schematic view of the ANKE magnetic spectrometer. It mainly consists
of three dipole magnets, an internal hydrogen cluster jet-target and three
detection systems (Pd-, Nd- and Fd-system). The red lines represent possible
tracks of positively charged particles and the blue lines of negatively
charged particles.} \label{ANKE}
\end{figure}

The data were taken with the magnetic spectrometer ANKE~\cite{Barsov:2001xj}
(cf.\ Fig.~\ref{ANKE} for a schematic representation of the setup), which is
part of an internal fixed-target experimental setup located at the COoler SYnchrotron
-- COSY of the Forschungszentrum J\"ulich. One of the main components of ANKE
is the magnetic system, with its three dipole magnets D1--D3. The accelerated
beam of unpolarized deuterons is deflected by the first dipole magnet D1
(cf.\ Fig.~\ref{ANKE}) into the target chamber, where the beam interacts with
the internal hydrogen cluster-jet target~\cite{Khoukaz:1999}.
The second dipole magnet D2 separates the ejectiles by their electric charge and momentum
into three different detection systems. The deuterons associated with $dp$
elastic scattering are deflected by D2 into the Forward (Fd) detection
system, which was the only element used in this experiment. The Fd was
designed and installed near the beam pipe to detect high-momentum particles.
Beam particles not interacting with the internal target are deflected by the
dipole magnets D2 and D3 back onto the nominal ring orbit. A special feature
of this magnetic spectrometer is the moveable D2 magnet, which can be shifted
perpendicular to the beam line. It is thus possible to optimise the
geometrical acceptance of the detection system for each reaction that one
would like to investigate.
The deuteron beam momentum range from $3120.17\;\textrm{MeV}/c$ to
$3204.16\;\textrm{MeV}/c$ was divided into 16 different fixed beam momenta
(cf.\ Table~\ref{tab:BeamMomenta}, originally for the determination of the $\eta$ meson mass \cite{Goslawski:2012dn}) using the supercycle mode of COSY. In each
supercycle it is possible to alternate between up to seven different beam
settings, each with a cycle length of $206\;\textrm{s}$. The beam momentum spread $\Delta
p_{\rm d}/p_{\rm d} < 6 \times 10^{-5}$ was determined using the spin
depolarisation technique~\cite{Goslawski:2009vf}.
\begin{table}[h]
\centering
\caption{Beam momenta $p_d$ for each supercycle and flattop in MeV/$c$.}
\scriptsize
\begin{tabular}{p{0.5cm}|p{0.7cm}p{0.7cm}p{0.7cm}p{0.7cm}p{0.7cm}p{0.7cm}p{0.7cm}}\hline
	& FT1  &FT2 &FT3  &FT4  &FT5  &FT6  &FT7  \\
	\hline
	SC1 & 3120.17& 3146.41 & 3148.45 & 3152.45 & 3158.71 & 3168.05 & 3177.51 \\
	SC2 & 3120.17& 3147.35 & 3150.42 & 3154.49 & 3162.78 & 3172.15 & 3184.87  \\
	SC3 & &  &  & 3157.48 & 3160.62 &  & 3204.16 \\
	\hline
\end{tabular}
\label{tab:BeamMomenta}
\end{table}

\section{Event Selection and Analysis}
\label{Analysis}

\begin{figure}[!h]
	\centering
	\includegraphics[width=1\linewidth]{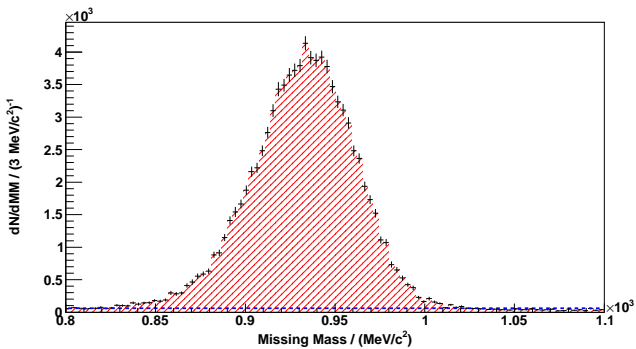}
	\caption{Missing-mass spectrum of the $dp\to dX$ reaction at $p_{d} =
		3120.17\;\textrm{MeV}/c$ for $0.08 < |t| < 0.09\;(\textrm{GeV}/c)^2$. The
		blue dashed line represents a constant background fit to the spectrum,
		excluding the $\pm3\sigma$ region around the peak.} \label{fig:MMElastic}
\end{figure}

As described above, deuterons originating from $dp$ elastic scattering are
deflected by D2 into the Forward detection system, which consists of one
multiwire drift chamber as well as two multiwire proportional chambers for
track reconstruction. In addition, two scintillator hodoscopes, comprised of
eight vertically aligned scintillator strips for the first and nine for the
second hodoscope, are used for particle identification using the energy-loss
information and time-of-flight measurements.
During the data taking a specific hardware trigger was included, which
required two coincident scintillator signals, one in each of the two Fd
hodoscopes. Due to the cross section for $dp$ elastic scattering being very
large, this hardware trigger is equipped with a pre-scaling factor of 1024 to
reduce the dead time of the data acquisition system.
On account of the small momentum transfer to the target proton, the
forward-going deuterons, whose tracks are reconstructed in the Forward
detection system, have momenta close to that of the beam. Since only
deuterons from elastic scattering have such a high momentum, the reaction can
be identified with no physical background from meson production.
Reconstructed particles with a momentum $p$ below about $p/p_{d} \approx
0.913$ are discarded to obtain a better signal-to-noise ratio.
In order to avoid uncertainties caused by small inhomogeneities of the
magnetic field at the edges of the D2 magnet, an additional cut in the $y$
hit position (with $y$ being the axis perpendicular to the COSY plane) of the
first multi-wire proportional chamber is required. Events with
$|y_{\textrm{{\scriptsize hit}}}| > 105\;\textrm{mm}$ are discarded.
For $dp$ elastic scattering the geometrical acceptance of the ANKE magnetic
spectrometer is limited to $0.06 < |t| < 0.31\;(\textrm{GeV}/c)^2$. However,
to avoid systematic edge effects, only events in the region $0.08 < |t| <
0.26\;(\textrm{GeV}/c)^2$ were analysed, with a bin width of $\Delta t =
0.01\;(\textrm{GeV}/c)^2$. The missing-mass analysis of
Fig.~\ref{fig:MMElastic} shows a prominent signal at the proton mass sitting
on top of a very small and seemingly constant background. 

A Gaussian fit to the peak was used to define its position and width and the region outside the
$\pm 3\sigma$ region was used to fit a constant background. After subtracting
this, the missing-mass spectra are integrated to obtain the number of $dp$
elastic scattering events for each of the 18 momentum transfer bins at all 16
different beam momenta.
The detector acceptance, which drops from 15\% to 7\% with increasing
momentum transfer, was determined using Monte Carlo simulations. These
simulations have to fulfil the same software cut criteria as the data, so
that the acceptance-corrected count yield can be determined for each beam
momentum setting. The resulting differential cross sections are presented in
Sec.~\ref{Results}.

\section{Theoretical calculation}
\label{Theory}

The theoretical calculation of the $pd$ elastic scattering cross section was
performed at four incident proton energies $T_p = 800$, $900$, $950$ and
$1000\;\textrm{MeV}$ within the refined Glauber
model~\cite{Platonova:2016xjq,Platonova:2010zz}. The differential cross
section is related to the amplitude $M$ as
\begin{equation}
\label{dsm}
\dd\sigma\! / \!\dd t = {\textstyle
\frac{1}{6}}{\rm Sp}\,\bigl(MM^+\bigr).
\end{equation}
The $pd$ amplitude $M$ in the Glauber approach contains two terms
corresponding to single and double scattering of the projectile with the
nucleons in the deuteron. These terms are expressed through the on-shell $NN$
amplitudes ($pp$ amplitude $M_p$ and $pn$ amplitude $M_n$) and the deuteron
wave function $\Psi_d$:
\begin{equation}
\label{msd} M({\bf q}) = M^{(s)}({\bf q}) + M^{(d)}({\bf q}),
\end{equation}
\begin{equation}
\label{ms}
M^{(s)}({\bf q}) = \int \dd^{3}r\, e^{i{\bf q}{\bf r}/2}
\Psi_d({\bf r}) \left[M_n({\bf q}) + M_p({\bf q})\right]
\Psi_d({\bf r}),
\end{equation}
\begin{equation}
\label{md}
M^{(d)}({\bf q}) = \frac{i}{4 \pi^{3/2}}\int \dd^{2}q' \int \dd^{3}r\, e^{i{\bf
	q'}{\bf r}} \Psi_d({\bf r}) \times
\end{equation}
\begin{equation*}
\Bigl[M_n({\bf q_2}) M_p({\bf q_1})
+ M_p({\bf q_2}) M_n({\bf q_1})
- M_c({\bf q_2}) M_c({\bf q_1})\Bigr] \Psi_d({\bf r}),
\end{equation*}
where ${\bf q}$ is the overall 3-momentum transfer (so that $t = -q^2$ in the
centre-of-mass system), while ${\bf q_1} = {\bf q}/2 - {\bf q'}$ and ${\bf
q_2} = {\bf q}/2 + {\bf q'}$ are the momenta transferred in collisions with
individual target nucleons, and $M_c({\bf q}) = M_n({\bf q}) - M_p({\bf q})$
is the amplitude of the charge-exchange process $pn \to np$.

When spin dependence is taken into account, the $NN$ amplitudes $M_n$, $M_p$
and the deuteron wave function $\Psi_d$ are non-commuting operators in the
three-nucleon spin space. They can be expanded into several independent terms
that are invariant under spatial rotations and space and time reflections,
and the coefficients of the expansions are, respectively, the $NN$ invariant
amplitudes (five for both $pp$ and $pn$ scattering) and $S$- and $D$-wave
components of the deuteron wave function. The $pd$ amplitude $M$ is also
expanded into $12$ independent terms. After undertaking some spin algebra and
integrating over the spatial coordinate, all the $pd$ invariant amplitudes
can be explicitly related to the $NN$ invariant amplitudes and the various
components of the deuteron form factor $S({\bf q}) = {\textstyle \int} \dd^3
r \, e^{i{\bf q}{\bf r}}|\Psi_d({\bf r})|^2$. The detailed derivation and the
final formulae of the refined Glauber model can be found in
Refs.~\cite{Platonova:2016xjq,Platonova:2010zz}.

The $NN$ invariant amplitudes at low momentum transfers are easily evaluated
from the centre-of-mass helicity amplitudes, which can be constructed from
empirical $NN$ phase shifts. For the present calculation, we used the phase
shifts of the latest PWA solution of the SAID group~\cite{Workman:2016ysf}.
There are, in fact, two PWA solutions published in
Ref.~\cite{Workman:2016ysf}, viz.\ the unweighted fit SM16 and the weighted
fit WF16. Unlike their earlier solution SP07~\cite{Arndt:2007qn}, both new SAID
solutions incorporate the recent high-precision COSY-ANKE
data~\cite{Mchedlishvili:2015iwa,Bagdasarian:2014mdj} on the near-forward
differential cross section ($1.0 \leq T_p \leq 2.8$~GeV) and analysing power
$A_y$ ($0.8 \leq T_p \leq 2.4$~GeV) in $pp$ elastic scattering and the
COSY-WASA data~\cite{Adlarson:2014pxj} on $A_y$ in $np$ scattering at $T_n =
1.135\;\textrm{GeV}$. However, by construction the WF16 solution describes
better the new COSY-ANKE results since the weights of these data have here
been enhanced.

The $NN$ partial-wave amplitudes obtained in the SM16, WF16 and SP07 solutions
begin to deviate significantly from each other only for $T_p \geq
1\;\textrm{GeV}$. We examined both new PWA solutions at $T_p =
900\;\textrm{MeV}$ and found the $pd$ differential cross section with WF16
input to be lower than that produced by SM16 by between 1\% and 3\% for $0.08
< |t| <0.26\;(\textrm{GeV}/c)^2$. This small difference is some measure of
the uncertainties arising from the input on-shell $NN$ amplitudes.

For three other energies ($T_p = 800$, $950$ and $1000\;\textrm{MeV}$) we
employed the WF16 $NN$ PWA solution and at $T_p = 1\;\textrm{GeV}$ we also
compared the results with those obtained with the SP07 input used in earlier
works~\cite{Platonova:2016xjq,Platonova:2010zz}. The changes ranged from 1\%
to 8\% in the momentum transfer interval $0.08 < |t| <
0.26\;(\textrm{GeV}/c)^2$.

Due to the rapid fall-off of the $NN$ amplitudes with momentum transfer, the
$pd$ predictions in the Glauber model are sensitive mainly to the long-range
behaviour of the deuteron wave function. We used the one derived from the
CD-Bonn $NN$-potential model~\cite{Machleidt:2000ge} but choosing a different
(but realistic) wave function would change the resulting $pd$ cross section
by not more than about $1$--$2$\%~\cite{Platonova:2010zz}.

The dependence of the $NN$ helicity amplitudes on the momentum transfer $q$,
as well as the dependence of the deuteron $S$- and $D$-wave functions on the
inter-nucleon distance $r$, were parameterised by convenient five-Gaussian
fits~\cite{Platonova:2016xjq,Platonova:2010zz}. The fitted $NN$ amplitudes
coincide with exact ones at momentum transfers $q < 0.7\;\textrm{GeV}/c$ and
the deuteron wave functions at distances $r < 20\;\textrm{fm}$. This
parametrisation allows us to perform the calculations fully analytically.

\section{Results}
\label{Results}

\begin{figure*}[!h]
\centering
\includegraphics[width=0.87\linewidth,trim = 0cm 0cm 0cm 1cm]{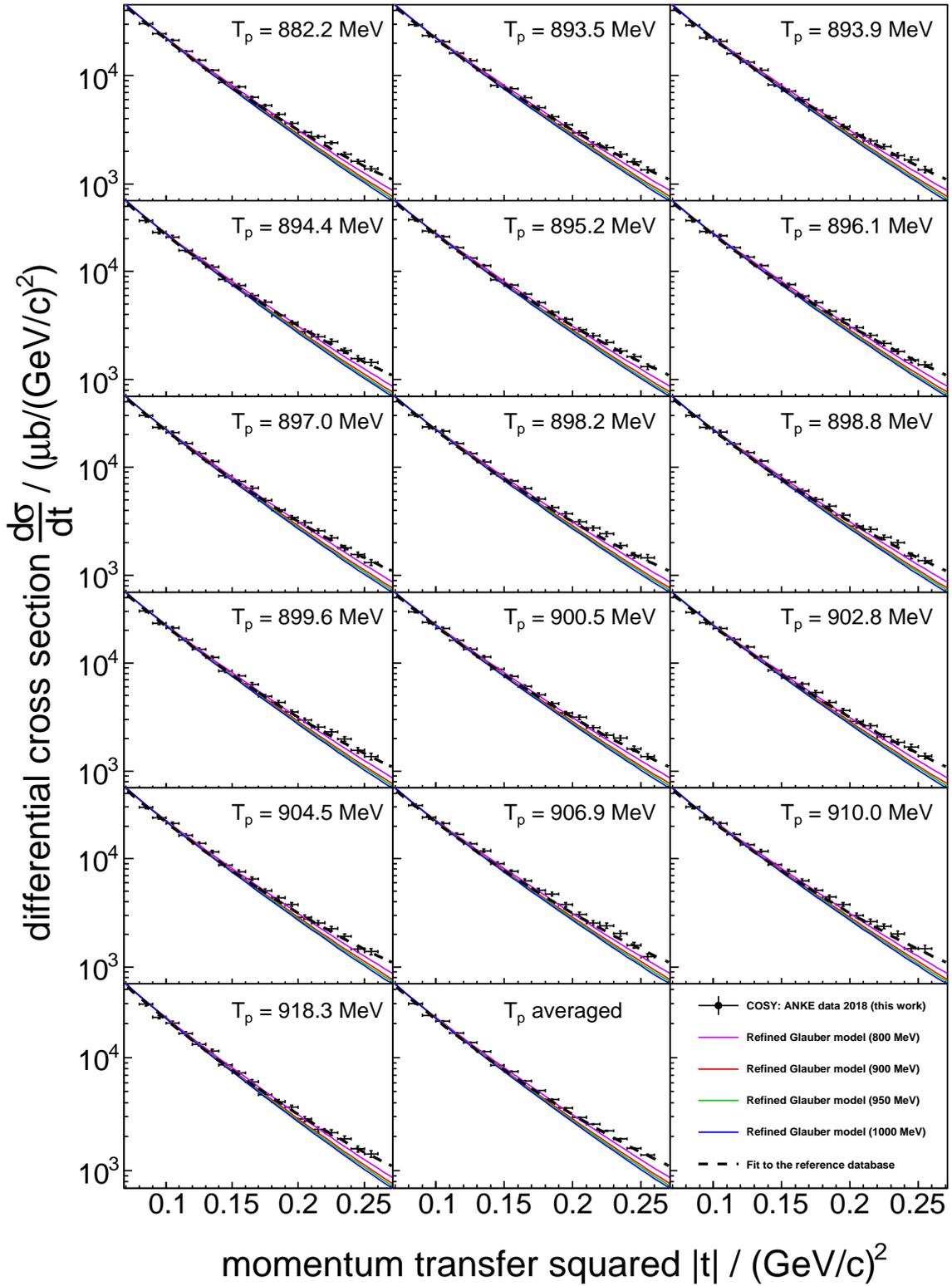}
\caption{Differential cross sections for deuteron-proton elastic scattering
for deuteron laboratory momenta between 3120.17 and
$3204.16\;\textrm{MeV}/c$. These are labeled in terms of the proton kinetic
energy for a deuteron target ($882.2 \leq T_p \leq 918.3\;\textrm{MeV}$).
Also shown is the average over the 16 available measurements. The purple
($T_p = 800\;\textrm{MeV}$), red ($T_p = 900\;\textrm{MeV}$), green ($T_p =
950\;\textrm{MeV}$), and blue ($T_p = 1000\;\textrm{MeV}$) lines represent
the refined Glauber model calculations (with the use of the SAID $NN$ PWA,
solution WF16~\cite{Workman:2016ysf}) and the dashed black line the fit to
the $dp$-elastic database from
\cite{Dalkhazhav:1969cma,Winkelmann:1980ca,Irom:1984wr,Velichko:1988ed,Guelmez:1991he}.}
\label{fig:DiffCrossSection}
\end{figure*}

The normalisation of the data presented here is obtained using the fit
\begin{equation}
\dd\sigma/\dd t = \exp(a + b|t| + c|t|^2)~\mu\textrm{b}/(\textrm{GeV}/c)^2
\end{equation}
in the momentum transfer range $0.05 < |t| < 0.4\;(\textrm{GeV}/c)^2$ to the
combined database from
Refs.~\cite{Dalkhazhav:1969cma,Winkelmann:1980ca,Irom:1984wr,Velichko:1988ed,Guelmez:1991he},
which led to the parameters $a = 12.45$, $b = -27.24\;(\textrm{GeV}/c)^{-2}$
and $c = 26.31\;(\textrm{GeV}/c)^{-4}$. To normalise the acceptance-corrected
counts at each beam momentum, both the fit to the reference database as well
as the numbers of counts are integrated over the momentum transfer range
$0.08 < |t| < 0.09\;(\textrm{GeV}/c)^2$. Assuming $\dd\sigma/\dd t$ is
independent of the beam momentum, the ratio between the two integrals defines
the scaling factor for each beam momentum that takes into account, e.g.,
different integrated luminosities. The differential cross sections thus
determined for all 16 beam momenta are shown in
Fig.~\ref{fig:DiffCrossSection}.

\begin{figure*}[t]
\centering
\includegraphics[width=1\linewidth]{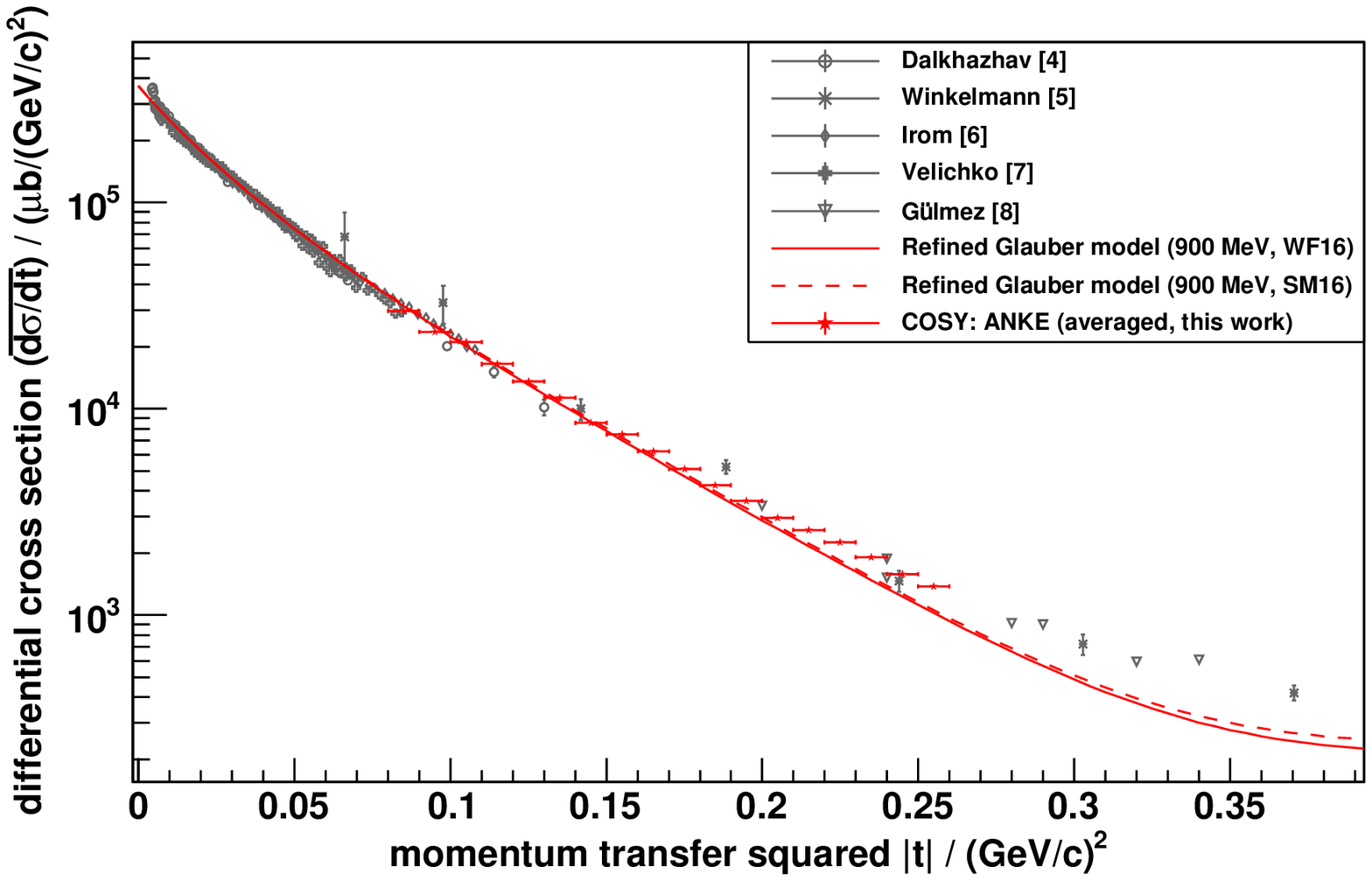}
\caption{Differential cross sections $\overline{d\sigma/dt}$ averaged over
the available 16 energies between $882.2\;\textrm{MeV} \leq T_p \leq
918.3\;\textrm{MeV}$ compared with the existing database
\cite{Dalkhazhav:1969cma,Winkelmann:1980ca,Irom:1984wr,Velichko:1988ed,Guelmez:1991he}
and the refined Glauber model calculation at $T_p = 900\;\textrm{MeV}$ (with
the use of the SAID $NN$ PWA, solutions WF16 and
SM16~\cite{Workman:2016ysf}).}
	\label{fig:NewWQDatabase}
\end{figure*}

The plots of differential cross sections at the 16 different beam momenta
shows that their shapes are independent of beam momentum over the available
momentum range. As a consequence, it is possible to evaluate the differential
cross section for each of the 18 momentum transfer bins averaged over the 16
energies (cf.\ Fig.~\ref{fig:DiffCrossSection}, Fig.~\ref{fig:NewWQDatabase},
and Table~\ref{tab:AveragedWQ}). The systematic uncertainties caused by,
e.g., the uncertainty in the angle calibration in the D2 magnet are
negligible compared to the statistical uncertainties that are presented in
Table~\ref{tab:AveragedWQ}.

\begin{table}[h]
\centering \caption{Differential cross section $\overline{\dd\sigma/\dd t}$
and statistical uncertainty of $dp$ elastic scattering averaged over all 16
different beam momenta.}
\begin{tabular}{ccccc}\\
	$|t|$ & $\overline{\dd\sigma/\dd t}$  & $\Delta \overline{\dd\sigma/\dd t}_{\textrm{\scriptsize stat}}$  \\
	$(\textrm{GeV}/c)^2$ & $\mu \textrm{b}/(\textrm{GeV}/c)^2$  & $\mu \textrm{b}/(\textrm{GeV}/c)^2$ \\
	\hline
	0.085 & 29898 & 193 \\
	0.095 & 23624 & 155 \\
	0.105 & 21014 & 140 \\
	0.115 & 16448 & 112 \\
	0.125 & 13562 & 95 \\
	0.135 & 11295 & 82 \\
	0.145 & 8546 & 65 \\
	0.155 & 7534 & 59 \\
	0.165 & 6212 & 51 \\
	0.175 & 5098 & 45 \\
	0.185 & 4264 & 39 \\
	0.195 & 3575 & 35 \\
	0.205 & 2963 & 31 \\
	0.215 & 2573 & 29 \\
	0.225 & 2249 & 26 \\
	0.235 & 1909 & 24 \\
	0.245 & 1575 & 21 \\
	0.255 & 1379 & 20 \\
	\hline
\end{tabular}
\label{tab:AveragedWQ}
\end{table}

\begin{figure}[!h]
	\centering
	\includegraphics[width=1\linewidth]{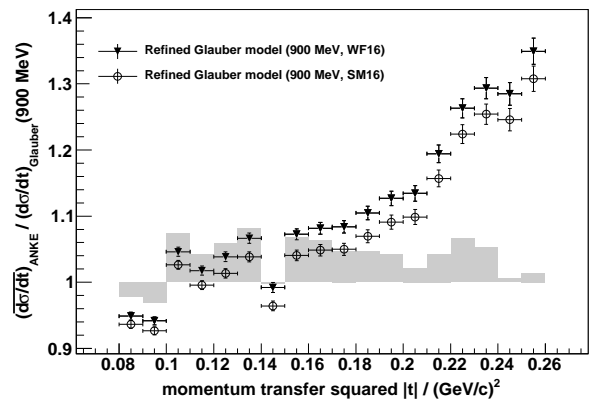}
	\caption{Ratio of our measured differential cross sections
		$\overline{d\sigma/dt}$ averaged over the available 16 energies to the
		refined Glauber model calculation at $T_p = 900\;\textrm{MeV}$ (with the use
		of the SAID $NN$ PWA, solutions WF16 and SM16~\cite{Workman:2016ysf}). The
		grey bars represent the ratio of the averaged differential cross sections to
		the fit to the reference database.}
	\label{fig:WQGlauberCompared}
\end{figure}

From the comparison of the results with the theoretical calculation at
$T_p=900\;\textrm{MeV}$ (see Figs.~\ref{fig:DiffCrossSection} and
\ref{fig:NewWQDatabase}), it is seen that the refined Glauber model describes
our data very well at low momentum transfers $0.08 < |t| <
0.2\;(\textrm{GeV}/c)^2$. It is also evident from
Fig.~\ref{fig:NewWQDatabase} that the refined Glauber model calculation
agrees similarly with the existing database for $|t| <
0.1\;(\textrm{GeV}/c)^2$. Fig.~\ref{fig:WQGlauberCompared} shows the ratio of
the averaged cross section determined in the present experiment to that
calculated within the refined Glauber model. The scatter of this ratio around
unity for $0.08 < |t| < 0.18\;(\textrm{GeV}/c)^2$ is consistent with the
scatter of experimental data around the smooth curve fitting the reference
database (see Fig.~\ref{fig:WQGlauberCompared}).

At the higher momentum transfers, the theoretical curve begins to deviate
from experiment and this is likely to be due to a failure of the
small-momentum-transfer approximations (account of only single and double
scattering, neglect of recoil, etc.) involved in the Glauber theory. On the
other hand, it was found in Ref.~\cite{Platonova:2016xjq} that at the lower
energies of $T_p = 250$ and $440\;\textrm{MeV}$ the refined Glauber model
calculations agree with the data on $pd$ elastic differential cross section
out to at least $|t| = 0.3\;(\textrm{GeV}/c)^2$, i.e., in the same region
where exact three-body Faddeev equations describe the data. However, the
accuracy of the Glauber model, which is a high-energy approximation to the
exact theory, should get better at higher collision energy.
The deviations noted here for $|t| > 0.2\;(\textrm{GeV}/c)^2$ might arise
from dynamical mechanisms that are not taken into account in either the
approximate (Glauber-like) or the exact (Faddeev-type) approach. For example,
there could be contributions from a three-nucleon ($3N$) force whose
importance rises with collision energy and momentum transfer. One
conventional $3N$-force, induced by two-pion exchange with an intermediate
$\Delta(1232)$-isobar excitation, is known to contribute to $pd$ large-angle
scattering at intermediate energies (see, e.g.,~\cite{Sekiguchi:2014vva}).
However, one might also consider three-body forces caused by the meson
exchange between the proton and the six-quark core of the deuteron (the
deuteron dibaryon) \cite{Kukulin:2003tk}. Indeed, at larger momentum transfers, the incident proton
probes shorter $NN$ distances in the deuteron, so that, the proton scattering
off the deuteron as a whole could occur with increasing probability. 
The preliminary results of taking the one-meson-exchange between the incident
proton and deuteron dibaryon into account in $pd$ elastic scattering have
shown this $3N$-force contribution to increase slightly the $pd$ differential
cross section already at moderate momentum transfers~\cite{Platonova:2012md}.
This interesting question clearly requires further investigation.
The calculations at different proton energies from 800 to 1000 MeV show a
gradual energy dependence of the $pd$ differential cross section (see Fig.
\ref{fig:DiffCrossSection}). The theoretical curves at four energies
intersect at around $|t| = 0.08\;(\textrm{GeV}/c)^2$ and then begin to
deviate from each other. The difference between the calculated cross sections
at $T_p = 800$ and 1000 MeV reaches 13\% at $|t| = 0.2\;(\textrm{GeV}/c)^2$.
The increasing slope of the curve implies that at these energies the
interaction radius in $pd$ (as well as $NN$) elastic scattering
effectively increases with energy. As a result, the forward diffraction peak
in the cross section becomes higher and narrower. This means that the $pd$
elastic cross section integrated over $0<|t| <0.2\;(\textrm{GeV}/c)^2$
increases slightly with energy (by $4$\% from $800$ to $1000\;\textrm{MeV}$),
though its part taken from $|t| = 0.08\;(\textrm{GeV}/c)^2$ (the lower limit
of the present experiment) decreases a little. Hence, whereas the $pd$
elastic cross section as a function of the momentum transfer squared is
usually assumed to be constant in the energy and momentum-transfer range
considered, the present model calculations reveal a slight energy dependence
of the magnitude and slope of the $pd$ elastic cross section. This result has
already been taken into account for normalisation of the recent COSY-WASA
experimental data on the $\eta$-meson production in $pd$
collisions~\cite{Adlarson:2018rgs}.

\section{Summary}
\label{Summary}

Due to its small number of active particles, deuteron-proton elastic
scattering at intermediate energies is well suited for the study of various
non-standard mechanisms of hadron interaction, such as the production of
nucleon isobars, dibaryon resonances, etc. However, even for $dp$ elastic
scattering, the experimental database is scarce at momentum transfers $|t| >
0.1\;(\textrm{GeV}/c)^2$. In this work, new precise measurements of the
differential cross sections for $dp$ elastic scattering at 16 equivalent
proton energies between $T_p = 882.2\;\textrm{MeV}$ and $T_p =
918.3\;\textrm{MeV}$ in the range $0.08 < |t| < 0.26\;(\textrm{GeV}/c)^2$
have been presented. Since the shapes of the differential cross sections were
found to be independent of beam momentum, it was possible to determine
precise average values over the whole momentum transfer range.

The experimental data at low momentum transfers $|t| <
0.2\;(\textrm{GeV}/c)^2$ are well described by the refined Glauber approach
at an average energy $T_p = 900\;\textrm{MeV}$. These calculations take full
account of spin degrees of freedom and use accurate input $NN$ amplitudes 
based on the most recent partial-wave analysis of the SAID
group~\cite{Workman:2016ysf}. The deviations of the theoretical predictions
from experimental data observed at the higher momentum transfers are likely to
be due to failure of the small-momentum-transfer approximations involved in
the Glauber model. These deviations might also reflect the missing
contributions of some dynamical mechanisms such as $3N$ forces.

The calculations at different energies, i.e., $T_p = 800$, $900$, $950$ and
$1000\;\textrm{MeV}$, show a slight energy dependence (increasing slope) in
the $pd$ elastic cross section as a function of momentum transfer squared
$|t|$. The predicted energy dependence may be trusted in the momentum
transfer region where the refined Glauber model describes the data. This
behaviour should be taken into account when using $pd$ elastic scattering for
the normalisation of other data. However, the energy dependence found in this
region is so weak that it cannot be identified in existing data. Very precise
measurements for at least two distinct energies (say, $T_p = 800$ and
$1000\;\textrm{MeV}$) would be needed to observe it.

In addition to the unpolarised differential cross sections, it would be
interesting to study the momentum transfer and energy behaviour of
polarisation observables (analysing powers, etc.), which can readily be
calculated within the refined Glauber model at the same energies $T_p =
800$--$1000\;\textrm{MeV}$. The theoretical predictions for such observables
will be presented in a forthcoming paper.

\section*{Acknowledgments}
\label{Acknow} We are grateful to other members of the ANKE collaboration and
the COSY crew for their work and the good experimental conditions during the
beam time. This work has been supported by the JCHP FEE and Russian
Foundation for Basic Research, grant No.~16-02-00265.


\bibliographystyle{model1a-num-names}
\bibliography{dpElasticPaper.bib}

\begin{thebibliography}{25}
\expandafter\ifx\csname natexlab\endcsname\relax\def\natexlab#1{#1}\fi
\providecommand{\url}[1]{\texttt{#1}}
\providecommand{\href}[2]{#2}
\providecommand{\path}[1]{#1}
\providecommand{\DOIprefix}{doi:}
\providecommand{\ArXivprefix}{arXiv:}
\providecommand{\URLprefix}{URL: }
\providecommand{\Pubmedprefix}{pmid:}
\providecommand{\doi}[1]{\href{http://dx.doi.org/#1}{\path{#1}}}
\providecommand{\Pubmed}[1]{\href{pmid:#1}{\path{#1}}}
\providecommand{\bibinfo}[2]{#2}
\ifx\xfnm\relax \def\xfnm[#1]{\unskip,\space#1}\fi
\bibitem[{Mersmann et~al.(2007)}]{Mersmann:2007gw}
\bibinfo{author}{T.~Mersmann}, et~al., \bibinfo{journal}{Phys. Rev. Lett.}
  \bibinfo{volume}{98} (\bibinfo{year}{2007}) \bibinfo{pages}{242301}.
  \DOIprefix\doi{10.1103/PhysRevLett.98.242301}.
  \href{http://arxiv.org/abs/nucl-ex/0701072}{\tt arXiv:nucl-ex/0701072}.
\bibitem[{Rausmann et~al.(2009)}]{Rausmann:2009dn}
\bibinfo{author}{T.~Rausmann}, et~al., \bibinfo{journal}{Phys. Rev.}
  \bibinfo{volume}{C80} (\bibinfo{year}{2009}) \bibinfo{pages}{017001}.
  \DOIprefix\doi{10.1103/PhysRevC.80.017001}.
  \href{http://arxiv.org/abs/0905.4595}{\tt arXiv:0905.4595}.
\bibitem[{Mielke et~al.(2014)}]{Mielke:2014xbu}
\bibinfo{author}{M.~Mielke}, et~al., \bibinfo{journal}{Eur. Phys. J.}
  \bibinfo{volume}{A50} (\bibinfo{year}{2014}) \bibinfo{pages}{102}.
  \DOIprefix\doi{10.1140/epja/i2014-14102-2}.
  \href{http://arxiv.org/abs/1404.2066}{\tt arXiv:1404.2066}.
\bibitem[{Dalkhazhav et~al.(1969)}]{Dalkhazhav:1969cma}
\bibinfo{author}{N.~Dalkhazhav}, et~al., \bibinfo{journal}{Sov. J. Nucl. Phys.}
  \bibinfo{volume}{8} (\bibinfo{year}{1969}) \bibinfo{pages}{196--202}.
  \bibinfo{note}{[Yad. Fiz. 8 (1968), 342]}.
\bibitem[{Winkelmann et~al.(1980)}]{Winkelmann:1980ca}
\bibinfo{author}{E.~Winkelmann}, et~al., \bibinfo{journal}{Phys. Rev.}
  \bibinfo{volume}{C21} (\bibinfo{year}{1980}) \bibinfo{pages}{2535--2541}.
  \DOIprefix\doi{10.1103/PhysRevC.21.2535}.
\bibitem[{Irom et~al.(1983)}]{Irom:1984wr}
\bibinfo{author}{F.~Irom}, et~al., \bibinfo{journal}{Phys. Rev.}
  \bibinfo{volume}{C28} (\bibinfo{year}{1983}) \bibinfo{pages}{2380--2385}.
  \DOIprefix\doi{10.1103/PhysRevC.28.2380}.
\bibitem[{Velichko et~al.(1988)}]{Velichko:1988ed}
\bibinfo{author}{G.~N. Velichko}, et~al., \bibinfo{journal}{Sov. J. Nucl.
  Phys.} \bibinfo{volume}{47} (\bibinfo{year}{1988}) \bibinfo{pages}{755--759}.
  \bibinfo{note}{[Yad. Fiz.47,1185(1988)]}.
\bibitem[{Guelmez et~al.(1991)}]{Guelmez:1991he}
\bibinfo{author}{E.~Guelmez}, et~al., \bibinfo{journal}{Phys. Rev.}
  \bibinfo{volume}{C43} (\bibinfo{year}{1991}) \bibinfo{pages}{2067--2076}.
  \DOIprefix\doi{10.1103/PhysRevC.43.2067}.
\bibitem[{Franco and Glauber(1966)}]{Franco:1965wi}
\bibinfo{author}{V.~Franco}, \bibinfo{author}{R.~J. Glauber},
  \bibinfo{journal}{Phys. Rev.} \bibinfo{volume}{142} (\bibinfo{year}{1966})
  \bibinfo{pages}{1195--1214}. \DOIprefix\doi{10.1103/PhysRev.142.1195}.
\bibitem[{Platonova and Kukulin(2010{\natexlab{a}})}]{Platonova:2016xjq}
\bibinfo{author}{M.~N. Platonova}, \bibinfo{author}{V.~I. Kukulin},
  \bibinfo{journal}{Phys. Rev.} \bibinfo{volume}{C81}
  (\bibinfo{year}{2010}{\natexlab{a}}) \bibinfo{pages}{014004}.
  \DOIprefix\doi{10.1103/PhysRevC.81.014004, 10.1103/PhysRevC.94.069902}.
  \href{http://arxiv.org/abs/1612.08694}{\tt arXiv:1612.08694},
  \bibinfo{note}{[Erratum: Phys. Rev.C94,no.6,069902(2016)]}.
\bibitem[{Platonova and Kukulin(2010{\natexlab{b}})}]{Platonova:2010zz}
\bibinfo{author}{M.~N. Platonova}, \bibinfo{author}{V.~I. Kukulin},
  \bibinfo{journal}{Phys. Atom. Nucl.} \bibinfo{volume}{73}
  (\bibinfo{year}{2010}{\natexlab{b}}) \bibinfo{pages}{86--106}.
  \DOIprefix\doi{10.1134/S1063778810010114}, \bibinfo{note}{[Yad.
  Fiz.73,90(2010)]}.
\bibitem[{Arndt et~al.(2007)Arndt, Briscoe, Strakovsky, and
  Workman}]{Arndt:2007qn}
\bibinfo{author}{R.~A. Arndt}, \bibinfo{author}{W.~J. Briscoe},
  \bibinfo{author}{I.~I. Strakovsky}, \bibinfo{author}{R.~L. Workman},
  \bibinfo{journal}{Phys. Rev.} \bibinfo{volume}{C76} (\bibinfo{year}{2007})
  \bibinfo{pages}{025209}. \DOIprefix\doi{10.1103/PhysRevC.76.025209}.
  \href{http://arxiv.org/abs/0706.2195}{\tt arXiv:0706.2195}.
\bibitem[{Workman et~al.(2016)Workman, Briscoe, and
  Strakovsky}]{Workman:2016ysf}
\bibinfo{author}{R.~L. Workman}, \bibinfo{author}{W.~J. Briscoe},
  \bibinfo{author}{I.~I. Strakovsky}, \bibinfo{journal}{Phys. Rev.}
  \bibinfo{volume}{C94} (\bibinfo{year}{2016}) \bibinfo{pages}{065203}.
  \DOIprefix\doi{10.1103/PhysRevC.94.065203}.
  \href{http://arxiv.org/abs/1609.01741}{\tt arXiv:1609.01741}.
\bibitem[{Mchedlishvili et~al.(2016)}]{Mchedlishvili:2015iwa}
\bibinfo{author}{D.~Mchedlishvili}, et~al., \bibinfo{journal}{Phys. Lett.}
  \bibinfo{volume}{B755} (\bibinfo{year}{2016}) \bibinfo{pages}{92--96}.
  \DOIprefix\doi{10.1016/j.physletb.2016.01.066}.
  \href{http://arxiv.org/abs/1510.06162}{\tt arXiv:1510.06162}.
\bibitem[{Bagdasarian et~al.(2014)}]{Bagdasarian:2014mdj}
\bibinfo{author}{Z.~Bagdasarian}, et~al., \bibinfo{journal}{Phys. Lett.}
  \bibinfo{volume}{B739} (\bibinfo{year}{2014}) \bibinfo{pages}{152--156}.
  \DOIprefix\doi{10.1016/j.physletb.2014.10.054}.
  \href{http://arxiv.org/abs/1409.8445}{\tt arXiv:1409.8445}.
\bibitem[{Adlarson et~al.(2014)}]{Adlarson:2014pxj}
\bibinfo{author}{P.~Adlarson}, et~al. (\bibinfo{collaboration}{WASA-at-COSY}),
  \bibinfo{journal}{Phys. Rev. Lett.} \bibinfo{volume}{112}
  (\bibinfo{year}{2014}) \bibinfo{pages}{202301}.
  \DOIprefix\doi{10.1103/PhysRevLett.112.202301}.
  \href{http://arxiv.org/abs/1402.6844}{\tt arXiv:1402.6844}.
\bibitem[{Barsov et~al.(2001)}]{Barsov:2001xj}
\bibinfo{author}{S.~Barsov}, et~al. (\bibinfo{collaboration}{ANKE}),
  \bibinfo{journal}{Nucl. Instrum. Meth.} \bibinfo{volume}{A462}
  (\bibinfo{year}{2001}) \bibinfo{pages}{364--381}.
  \DOIprefix\doi{10.1016/S0168-9002(00)01147-5}.
\bibitem[{Khoukaz et~al.(1999)}]{Khoukaz:1999}
\bibinfo{author}{A.~Khoukaz}, et~al., \bibinfo{journal}{Eur. Phys. J.}
  \bibinfo{volume}{D5} (\bibinfo{year}{1999}) \bibinfo{pages}{275}.
\bibitem[{Goslawski et~al.(2012)}]{Goslawski:2012dn}
\bibinfo{author}{P.~Goslawski}, et~al., \bibinfo{journal}{Phys. Rev.}
  \bibinfo{volume}{D85} (\bibinfo{year}{2012}) \bibinfo{pages}{112011}.
  \DOIprefix\doi{10.1103/PhysRevD.85.112011}.
  \href{http://arxiv.org/abs/1204.3520}{\tt arXiv:1204.3520}.
\bibitem[{Goslawski et~al.(2010)}]{Goslawski:2009vf}
\bibinfo{author}{P.~Goslawski}, et~al., \bibinfo{journal}{Phys. Rev. ST Accel.
  Beams} \bibinfo{volume}{13} (\bibinfo{year}{2010}) \bibinfo{pages}{022803}.
  \DOIprefix\doi{10.1103/PhysRevSTAB.13.022803}.
  \href{http://arxiv.org/abs/0908.3103}{\tt arXiv:0908.3103}.
\bibitem[{Machleidt(2001)}]{Machleidt:2000ge}
\bibinfo{author}{R.~Machleidt}, \bibinfo{journal}{Phys. Rev.}
  \bibinfo{volume}{C63} (\bibinfo{year}{2001}) \bibinfo{pages}{024001}.
  \DOIprefix\doi{10.1103/PhysRevC.63.024001}.
  \href{http://arxiv.org/abs/nucl-th/0006014}{\tt arXiv:nucl-th/0006014}.
\bibitem[{Sekiguchi et~al.(2014)}]{Sekiguchi:2014vva}
\bibinfo{author}{K.~Sekiguchi}, et~al., \bibinfo{journal}{Phys. Rev.}
  \bibinfo{volume}{C89} (\bibinfo{year}{2014}) \bibinfo{pages}{064007}.
  \DOIprefix\doi{10.1103/PhysRevC.89.064007}.
\bibitem[{Kukulin et~al.(2004)Kukulin, Pomerantsev, Kaskulov, and
  Faessler}]{Kukulin:2003tk}
\bibinfo{author}{V.~I. Kukulin}, \bibinfo{author}{V.~N. Pomerantsev},
  \bibinfo{author}{M.~Kaskulov}, \bibinfo{author}{A.~Faessler},
  \bibinfo{journal}{J. Phys.} \bibinfo{volume}{G30} (\bibinfo{year}{2004})
  \bibinfo{pages}{287--308}. \DOIprefix\doi{10.1088/0954-3899/30/3/005}.
  \href{http://arxiv.org/abs/nucl-th/0308059}{\tt arXiv:nucl-th/0308059}.
\bibitem[{Platonova and Kukulin(2012)}]{Platonova:2012md}
\bibinfo{author}{M.~N. Platonova}, \bibinfo{author}{V.~I. Kukulin},
  \bibinfo{journal}{J. Phys. Conf. Ser.} \bibinfo{volume}{381}
  (\bibinfo{year}{2012}) \bibinfo{pages}{012110}.
  \DOIprefix\doi{10.1088/1742-6596/381/1/012110}.
\bibitem[{Adlarson et~al.(2018)}]{Adlarson:2018rgs}
\bibinfo{author}{P.~Adlarson}, et~al. (\bibinfo{collaboration}{WASA-at-COSY})
  (\bibinfo{year}{2018}). \href{http://arxiv.org/abs/1801.06671}{\tt
  arXiv:1801.06671}.

\end{thebibliography}

\end{document}